\pdfoutput=1
\RequirePackage{ifpdf}
\ifpdf % We are running pdfTeX in pdf mode
\documentclass[pdftex]{sigma}
\else
\documentclass{sigma}
\fi

{ \theoremstyle{definition}
\newtheorem*{Remark}{Remark}
 }

\begin{document}

\allowdisplaybreaks

\renewcommand{\PaperNumber}{045}

\FirstPageHeading

\ShortArticleName{B\"acklund--Darboux Transformations and Discretizations of Super KdV Equation}

\ArticleName{B\"acklund--Darboux Transformations\\
and Discretizations of Super KdV Equation}

\Author{Ling-Ling XUE and Qing Ping LIU}

\AuthorNameForHeading{L.L.~Xue and Q.P.~Liu}

\Address{Department of Mathematics, China University of Mining and Technology,\\
Beijing 100083, P.~R.~China}
\Email{\href{mailto:xue0000@126.com}{xue0000@126.com}, \href{mailto:qpl@cumtb.edu.cn}{qpl@cumtb.edu.cn}}

\ArticleDates{Received January 02, 2014, in f\/inal form April 10, 2014; Published online April 17, 2014}

\Abstract{For a~generalized super KdV equation, three Darboux transformations and the corresponding B\"acklund
transformations are constructed.
The compatibility of these Darboux transformations leads to three discrete systems and their Lax representations.
The reduction of one of the B\"acklund--Darboux transformations and the corresponding discrete system are considered for
Kupershmidt's super KdV equation.
When all the odd variables vanish, a~nonlinear superposition formula is obtained for Levi's B\"acklund transformation
for the KdV equation.}


\Classification{35Q53; 37K10; 35A30}

\section{Introduction}

It is well known that the modern theory of integrable systems or soliton theory begins with the study of the celebrated
KdV equation by Kruskal and his collaborators~\cite{kruskal}.
Various types of extensions of this equation exist in literature (see~\cite{ablowitz} for example) and one of them is
the super extensions.
The f\/irst such extension was proposed by Kupershmidt~\cite{BAK}, which reads as
\begin{gather}
u_t=u_{xxx}-6u u_x+12\xi_{xx}\xi,\nonumber
\\
\xi_t=4\xi_{xxx}-6u\xi_x-3u_x\xi,\label{eq1}
\end{gather}
where subscripts denote partial derivatives, $t$ and $x$ are the temporal variable and spatial variable respectively.
$u$~is a~bosonic (even or commuting) variable and $\xi$ is a~fermionic (odd or anti-commuting) variable which fulf\/ill
\begin{gather*}
\xi^{(i)}\xi^{(j)}=-\xi^{(j)}\xi^{(i)},
\qquad
u^{(i)}u^{(j)}=u^{(j)}u^{(i)},
\qquad
\xi^{(i)}u^{(j)}=u^{(j)}\xi^{(i)},
\end{gather*}
where $(\,)^{(i)}={\partial_x}^{i}(\,)$.
For $\xi=0$,~\eqref{eq1} becomes the KdV equation.
Like the KdV equation itself, the super KdV equation~\eqref{eq1}, being a~bi-Hamiltonian system and possessing Lax
representation, is integrable in the conventional sense.

A dif\/ferent super KdV equation was proposed slightly later by Manin and Radul~\cite{mr} in their study of the
supersymmetric KP hierarchy.
This system, being the simplest and most important reduction of the supersymmetric KP hierarchy, reads as
\begin{gather}
u_t=u_{xxx}+6u u_x+3\xi_{xx}\xi,\nonumber
\\
\xi_t=\xi_{xxx}+3(u\xi)_x.\label{eq2}
\end{gather}
Even though the above systems~\eqref{eq1} and~\eqref{eq2} are similar in appearance, they are very dif\/ferent.
In fact, as observed by Mathieu~\cite{ma_PLB, ma_JMP}, the latter is invariant under the following transformation
\begin{gather*}
\widetilde{u}=u+\epsilon \xi_x,
\qquad
\widetilde{\xi}=\xi+\epsilon u,
\end{gather*}
where $\epsilon$ is a~fermionic parameter.
Then one may introduce a~new independent fermionic va\-riab\-le~$\theta$ and super f\/ield
$\alpha=\xi+\theta u$,
together with the corresponding super derivative ${\mathcal D}=\partial_\theta+\theta\partial_x$.
In this way, the system~\eqref{eq2} may be reformulated as a~single equation
\begin{gather*}
\alpha_t=\alpha_{xxx}+3 (\alpha{\mathcal D}\alpha )_{x}.
\end{gather*}
For this reason, the system~\eqref{eq1} is often referred as the super or fermionic KdV equation, while the
system~\eqref{eq2} is known as the supersymmetric KdV equation.

Nowadays, discrete integrable systems are very hot topic in the soliton theory, and to construct the discrete versions
of the non-commuting extensions of integrable equations is very interesting.
Most recently, Grahovski and Mikhailov~\cite{sasha} proposed integrable discretizations for a~class of nonlinear
Schr\"odinger equations on Grassmann algebras.
Also, with Levi we succeeded in discretizing the supersymmetric KdV equation~\eqref{eq2} and both semi-discrete and
fully discrete supersymmetric KdV equations are given~\cite{xll}.
The aim of this paper is to study Kupershmidt's super KdV equation~\eqref{eq1} in the same spirits.
In the following discussion, we will assume that $u$ and $\xi$ depend on not only continuous variables $x$ and $t$, but
also are functions of integer-valued variables $n$ and $m$.
The subscripts $_{[1]}$ and $_{[2]}$ used in the following denote the shifts of the discrete variables, for example,
$\xi_{[1]}=\xi(x,t,n+1,m)$, $\xi_{[2]}=\xi(x,t,n,m+1)$.

The outline of this paper is as follows.
In Section~\ref{section2}, we recall a~generalized super KdV system and its Lax representation.
In Section~\ref{section3}, three dif\/ferent Darboux and B\"{a}cklund transformations are worked out for the generalized super KdV system.
Then in Section~\ref{section4}, we employ these transformations to construct discrete integrable super systems and the relevant
reductions are discussed.
Using two kinds of elementary Darboux transformations, we obtain two dif\/ference-dif\/ference equations.
And by a~pair of binary Darboux transformations, we get a~dif\/ferential-dif\/ference equation.
The f\/inal section summarizes the results.

\section{A generalized super KdV system}\label{section2}

We aim to construct Darboux and B\"acklund transformations for the super KdV equation~\eqref{eq1}.
To this end, our strategy is to consider a~more general super system
\begin{gather}
u_t=u_{xxx}-6u u_x+6\xi_{xx}\eta+6\eta_{xx}\xi,
\nonumber\\
\xi_t=4\xi_{xxx}-6u\xi_x-3u_x\xi,
\nonumber\\
\eta_t=4\eta_{xxx}-6u\eta_x-3u_x\eta,\label{eq3}
\end{gather}
where $u=u(x,t)$ is a~bosonic variable, $\xi=\xi(x,t)$ and $\eta=\eta(x,t)$ are fermionic variables.
To the best of our knowledge, above system was studied f\/irst by Holod and Pakuliak~\cite{hp}.
The associated spectral problem is
\begin{gather}
L\psi=\lambda\psi,
\qquad
L=\partial_{x}^2-u-\xi\partial_{x}^{-1}\eta,
\nonumber\\
\psi_t=P \psi,
\qquad
P=4(L^{\frac{3}{2}})_{+}=4\partial_{x}^3-6 u \partial_{x}-3 u_x-6\xi\eta.\label{eq4}
\end{gather}
Introducing $\sigma_x=\eta\psi$ and $\chi=(\psi, \psi_x, \sigma)^{\text{T}}$, then we may rewrite~\eqref{eq4} in
matrix form, that is,
\begin{gather}
\chi_{x}={\mathcal L}\chi,
\qquad
{\mathcal L}= \left(
\begin{matrix} 0 & 1 & 0
\\
\lambda +u& 0 & \xi
\\
\eta & 0 & 0
\end{matrix}
\right),\label{LAXL_kuper}
\\
%\label{LAXP_kuper}
\chi_{t}={\mathcal P}\chi,
\qquad
{\mathcal P}= \left(
\begin{matrix}
u_x-2\xi\eta & 4\lambda-2u& 4\xi_x
\\
Z& -u_x-2\xi\eta & 4\xi_{xx}+(4\lambda-2 u)\xi
\\
4\eta_{xx}+(4\lambda-2 u)\eta& -4\eta_x & -4\xi\eta
\end{matrix}
\right),\nonumber
\end{gather}
where
\begin{gather*}
Z\equiv u_{xx}+(\lambda+u)(4\lambda-2u)+2(\xi_x\eta-\xi\eta_x).
\end{gather*}
A direct calculation shows that the Lax equation
\begin{gather*}
L_t=[P,L],
\end{gather*}
or the zero curvature condition
\begin{gather*}
{\mathcal L}_t-{\mathcal P}_x=[{\mathcal P},{\mathcal L}]
\end{gather*}
gives the generalized super KdV system~\eqref{eq3}.

\begin{Remark}
For $\xi=\eta$,~\eqref{eq3} reduces to Kupershmidt's super KdV equation~\eqref{eq1}.
\end{Remark}
\begin{Remark}
For all fermionic variables disappear,~\eqref{eq3} reduces to the KdV equation with linear spectral problem
\begin{gather*}
\chi_{x}={\mathcal L}\chi,
\qquad\!\!
{\mathcal L}= \left(
\begin{matrix} 0 & 1
\\
\lambda +u\!\!& 0
\end{matrix}
\right),
\qquad
\chi_{t}={\mathcal P}\chi,
\qquad\!\!
{\mathcal P}= \left(
\begin{matrix} u_x & 4\lambda-2u
\\
u_{xx}\!+(\lambda+u)(4\lambda-2u)\!\!& -u_x
\end{matrix}
\right).
\end{gather*}
\end{Remark}

In the following, we will use~\eqref{LAXL_kuper} to construct B\"acklund and Darboux transformations.

\section{Darboux and B\"{a}cklund transformations}\label{section3}

Now we manage to construct Darboux and B\"acklund transformations for the generalized super KdV system~\eqref{eq3}.
For convenience, we introduce the potentials $w$ and $w_{[i]}$ such that $u=w_x$, $u_{[i]}={w_{[i]}}_x$, and def\/ine
$v_i=w_{[i]}-w$ for $i=1, 2$.
Also, suppose that $\chi_{[0]}=(\psi_{[0]}, {\psi_{[0]}}_x, \sigma_{[0]})^{\text{T}}$ is a~solution
of~\eqref{LAXL_kuper} for $\lambda=p_1$, then we f\/ind three Darboux transformations and their corresponding B\"acklund
transformations, which are listed below.

 {\bf Case 1.} Def\/ine
\begin{gather*}
v_1\equiv -2(\ln{\psi_{[0]}})_x,
\qquad
\xi_{[1]}\equiv \xi_x+\frac{1}{2}v_1\xi,
\qquad
\eta_{[1]} \equiv -\frac{\sigma_{[0]}}{\psi_{[0]}}
%\label{field}
\end{gather*}
and
\begin{gather}
\label{lax1_kuper}
\chi_{[1]}\equiv{\mathcal W}\chi,
\qquad
{\mathcal W}= \left(
\begin{matrix}
\frac{1}{2}v_1 & 1 & 0
\vspace{1mm}\\
\lambda-p_1+\frac{1}{4}{v_1}^2+\xi\eta_{[1]}& \frac{1}{2}v_1 & \xi
\vspace{1mm}\\
\eta_{[1]} & 0 & 1
\end{matrix}
\right),
\end{gather}
then $\chi_{[1]}$ satisf\/ies
\begin{gather}
\label{gauge_1}
{\chi_{[1]}}_x={\mathcal L}_{[1]}\chi_{[1]},
\qquad
{\mathcal L}_{[1]}= \left(
\begin{matrix}
0 & 1 & 0
\\
\lambda +u_{[1]}& 0 & \xi_{[1]}
\\
\eta_{[1]} & 0 & 0
\end{matrix}
\right).
\end{gather}
The compatibility of the two linear systems~\eqref{lax1_kuper} and~\eqref{gauge_1} yields
\begin{gather*}
{\mathcal W}_x+{\mathcal W}{\mathcal L}-{\mathcal L}_{[1]}{\mathcal W}=0,
\end{gather*}
which leads to a~B\"{a}cklund transformation
\begin{gather}
\xi_{[1]}=\xi_x+\frac{1}{2}v_1\xi,
\qquad
\eta=-{\eta_{[1]}}_x+\frac{1}{2}v_1\eta_{[1]},
\qquad
{w_{[1]}}_x=-w_{x}+\frac{1}{2}{v_1}^2-2 p_1+2\xi\eta_{[1]}.\label{eq9}
\end{gather}
\begin{Remark}
During the 5th International Workshop on Nonlinear Mathematical Physics and the 12th National
Conference on Integrable Systems, held in Hangzhou last summer, we learnt that professor R.G.~Zhou also considered such
Darboux transformation~\cite{zhou}.
\end{Remark}

 {\bf Case 2.} Def\/ine
\begin{gather*}
{\overline\sigma_{[0]}}_x=\xi\psi_{[0]},
\qquad
v_1\equiv -2(\ln{\psi_{[0]}})_x+2\frac{\overline\sigma_{[0]}\sigma_{[0]}}{\psi_{[0]}^2},
\qquad
\xi_{[1]}\equiv \frac{\overline\sigma_{[0]}}{\psi_{[0]}},
\qquad
\eta_{[1]}\equiv-{\eta}_x-\frac{1}{2}v_1\eta
%\label{field2}
\end{gather*}
and
\begin{gather}
\label{lax2_kuper}
\chi_{[1]}\equiv{\mathcal W}\chi,
\qquad
{\mathcal W}= \left(
\begin{matrix}
\frac{1}{2}v_1 & 1 & -\xi_{[1]}
\vspace{1mm}\\
 \lambda- p_{1}+\frac{1}{4}{v_1}^2 & \frac{1}{2}v_1 & - \frac{1}{2}v_1 \xi_{[1]}
\vspace{1mm}\\
-\frac{1}{2}v_1\eta & -\eta & \lambda- p_{1} -\xi_{[1]}\eta
\end{matrix}
\right),
\end{gather}
then $\chi_{[1]}$ satisf\/ies
\begin{gather}
\label{gauge_2}
{\chi_{[1]}}_x={\mathcal L}_{[1]}\chi_{[1]}.
\end{gather}
The compatibility of the linear systems~\eqref{lax2_kuper} and~\eqref{gauge_2} supplies
\begin{gather*}
{\mathcal W}_x+{\mathcal W}{\mathcal L}-{\mathcal L}_{[1]}{\mathcal W}=0,
\end{gather*}
which gives the following B\"{a}cklund transformation
\begin{gather}
\xi={\xi_{[1]}}_x-\frac{1}{2}v_1\xi_{[1]},
\qquad
\eta_{[1]}=-{\eta}_x-\frac{1}{2}v_1\eta,
\qquad
{w_{[1]}}_x=-w_{x}+\frac{1}{2}{v_1}^2-2 p_1+2\xi_{[1]}\eta.\label{eq12}
\end{gather}

\begin{Remark}
For all fermionic variables vanish, it is easy to see that above Darboux transformations and B\"acklund transformations
reduce to the well-known results for the KdV equation.
\end{Remark}

\begin{Remark}
We observe that there exists a~symmetry between B\"acklund transformation~\eqref{eq9} and B\"ack\-lund
transformation~\eqref{eq12}.
In fact, if we make the following replacements:
\begin{gather*}
\xi\leftrightarrow \xi_{[1]},
\qquad
\eta\leftrightarrow \eta_{[1]},
\qquad
w\leftrightarrow w_{[1]},
\end{gather*}
then~\eqref{eq9} becomes~\eqref{eq12}.
\end{Remark}

{\bf Case 3.} Def\/ine
\begin{gather*}
{\overline\sigma_{[0]}}_x=\xi\psi_{[0]},
\qquad
F_x = \psi_{[0]}^2+\frac{F}{\psi_{[0]}^2}\overline{\sigma}_{[0]}{\sigma}_{[0]},
\\
v_1\equiv -2\frac{\psi_{[0]}^2}{F},
\qquad
\xi_{[1]}\equiv \xi-\frac{\psi_{[0]}}{F}\overline{\sigma}_{[0]},
\qquad
\eta_{[1]}\equiv \eta-\frac{\psi_{[0]}}{F}\sigma_{[0]}
%\label{field3}
\end{gather*}
and
\begin{gather*}
\chi_{[1]}\equiv{\mathcal W}\chi,
\\
{\mathcal W}= \left(
\begin{matrix} \lambda+A& \frac{1}{2} v_1 & \xi -\xi_{[1]}
\\
\frac{1}{2}\lambda v_1+B & \lambda+A +\frac{{v_1}_x}{2} & -\left(\frac{v_1}{4}+\frac{{v_1}_x}{2 v_1}
\right)(\xi_{[1]}-\xi)
\\
\left(\frac{v_1}{4}-\frac{{v_1}_x}{2 v_1}\right)(\eta_{[1]}-\eta)& \eta_{[1]}-\eta &
\lambda-p_1+\frac{2}{v_1}(\xi_{[1]}-\xi)(\eta_{[1]}-\eta)
\end{matrix}
\right)
\end{gather*}
with
\begin{gather*}
\begin{split}
& A\equiv -p_1-\frac{{v_1}_x}{4}+\frac{{v_1}^2}{8}+\frac{1}{v_1} (\xi_{[1]}-\xi ) (\eta_{[1]}-\eta ),
\\
& B\equiv-\frac{{{v_1}_{x}}^2}{8 v_1}+\frac{{v_1}^3}{32}-\frac{p_1 }{2} v_1 + \frac{1}{2}
 (\xi_{[1]}-\xi ) (\eta_{[1]}-\eta ),
 \end{split}
\end{gather*}
then $\chi_{[1]}$ solves
\begin{gather*}
{\chi_{[1]}}_x={\mathcal L}_{[1]}\chi_{[1]}.
\end{gather*}
Similarly, the compatibility of the above linear systems leads to
\begin{gather*}
{\mathcal W}_x+{\mathcal W}{\mathcal L}-{\mathcal L}_{[1]}{\mathcal W}=0,
\end{gather*}
or the B\"{a}cklund transformation
\begin{gather}
{\xi_{[1]}}_x = \xi_x +\frac{v_1}{2}\xi +\left(\frac{v_1}{4}+\frac{{v_1}_x}{2 v_1}\right)  (\xi_{[1]}-\xi ),
\nonumber\\
{\eta_{[1]}}_x= \eta_x+\frac{v_1}{2}\eta+\left(\frac{v_1}{4}+\frac{{v_1}_x}{2 v_1}\right)  (\eta_{[1]}-\eta ),
\nonumber\\
{w_{[1]}}_{xx}={w}_{xx}+v_1  ( {v_{1}}_x+2w_x  )+ \frac{{{v_1}_{x}}^2}{2 v_1}-\frac{{v_1}^3}{8}+2 p_1 v_1
+2 ( \eta\xi_{[1]}+\xi\eta_{[1]} ).\label{eq13}
\end{gather}

It is remarked that in the last case if the fermionic variables $\xi$ and $\eta$ vanish, we recover a~Darboux
transformation and related B\"{a}cklund transformation for the KdV equation, which are nothing but the ones obtained by
Levi~\cite{levi}.
Such a~Darboux transformation is also known as binary Darboux transformation in literature~\cite{matveev}.
Explicitly, the related B\"acklund transformation~\cite{levi} reads as
\begin{gather}
\label{levi_bt}
{w_{[1]}}_{xx}={w}_{xx}+v_1({v_{1}}_x+2w_x)+ \frac{{{v_1}_{x}}^2}{2 v_1}-\frac{{v_1}^3}{8}+2 p_1 v_1.
\end{gather}

Darboux transformations of binary type may be regarded as the composition of elementary Darboux
transformations~\cite{nimmo} (see also~\cite{cieslinski}).
Thus it is natural to expect that Levi's B\"{a}cklund transformation~\eqref{levi_bt} is also the composition of
elementary B\"{a}cklund transformations and this is indeed the case.
To see this we consider two copies of elementary B\"{a}cklund transformation for the KdV equation, namely
\begin{gather*}
(w_{[1]}+\bar{w})_x=\frac{1}{2} (w_{[1]}-\bar{w} )^2-2p_1,
\qquad
(w+\bar{w})_x=\frac{1}{2} (\bar{w}-w )^2-2p_2,
\end{gather*}
then, eliminating $\bar{w}$ leads to
\begin{gather*}
{w_{[1]}}_{xx}={w}_{xx}+v_1({v_{1}}_x+2w_x)+ \frac{{{v_1}_{x}}^2}{2 v_1}-\frac{{v_1}^3}{8}+(p_1+p_2)
v_1-\frac{2}{v_1}{(p_1-p_2)^2},
\end{gather*}
which reduces to~\eqref{levi_bt} if $p_1=p_2$.
Similar idea works for the super case and we can obtain the B\"{a}cklund transformation~\eqref{eq13} by the
superposition of the elementary B\"{a}cklund transformations~\eqref{eq9} or~\eqref{eq12}.
Thus, elementary Darboux/B\"{a}cklund transformations are more fundamental and binary Darboux/B\"{a}cklund
transformations are more involved.
The point is that while it is not clear how to reduce the Darboux/B\"acklund transformations of the Cases~1 and~2 to
Kupershmidt's super KdV equation~\eqref{eq1}, the reduction is feasible and easy to implement for the last case.
Indeed, for $\xi=\eta$, def\/ine
\begin{gather*}
F_x = \psi_{[0]}^2,
\qquad
v_1\equiv-2(\ln{F})_{x},
\qquad
\xi_{[1]}\equiv \xi-\frac{\psi_{[0]}}{F}{\sigma}_{[0]}
\end{gather*}
and
\begin{gather*}
\chi_{[1]}\equiv{\mathcal W}\chi,
\qquad
{\mathcal W}= \left(
\begin{matrix} \lambda+A& \frac{1}{2} v_1 & \xi -\xi_{[1]}
\\
\frac{1}{2}\lambda v_1+B & \lambda+A +\frac{{v_1}_x}{2} & -\left(\frac{v_1}{4}+\frac{{v_1}_x}{2 v_1}
\right)(\xi_{[1]}-\xi)
\\
\left(\frac{v_1}{4}-\frac{{v_1}_x}{2 v_1}\right)(\xi_{[1]}-\xi)& \xi_{[1]}-\xi & \lambda-p_1
\end{matrix}
\right)
\end{gather*}
with
\begin{gather*}
A\equiv -p_1-\frac{{v_1}_x}{4}+\frac{{v_1}^2}{8},
\qquad
B\equiv-\frac{{{v_1}_{x}}^2}{8 v_1}+\frac{{v_1}^3}{32}-\frac{p_1 }{2} v_1,
\end{gather*}
then it is straightforward to check that $\chi_{[1]}$ satisf\/ies
\begin{gather*}
{\chi_{[1]}}_x={\mathcal L}_{[1]}\chi_{[1]}.
\end{gather*}
The corresponding B\"acklund transformation is
\begin{gather}
{\xi_{[1]}}_x = \xi_x +\frac{v_1}{2}\xi +\left(\frac{v_1}{4}+\frac{{v_1}_x}{2 v_1}\right)  (\xi_{[1]}-\xi ),
\nonumber\\
{w_{[1]}}_{xx}={w}_{xx}+v_1({v_{1}}_x+2w_x)+ \frac{{{v_1}_{x}}^2}{2 v_1}-\frac{{v_1}^3}{8}+2 p_1 v_1+4\xi\xi_{[1]}.\label{eq15}
\end{gather}
Thus, we obtain a~B\"{a}cklund transformation for Kupershmidt's super KdV equation~\eqref{eq1}.

\section{Discrete systems}\label{section4}

Integrable discretizations have been studied extensively since the seventies of last century and various approaches have
been proposed (see~\cite{Nijihoff,Suris}).
Among them, the method, based on Darboux and B\"{a}cklund transformations, which f\/irst appeared
in~\cite{levi1981,levi1980}, has been proved to be very fruitful.
This idea is also applicable to super integrable systems and supersymmetric integrable systems~\cite{sasha,xll}.
We now adopt this idea for the super KdV equation~\eqref{eq3} and its reduction~-- Kupershmidt's super KdV equation to
construct their discrete counterparts.

We start our consideration with Darboux transformations presented in the f\/irst two cases of last section.

{\bf  Case A.} Consider a~pair of Darboux transformations which are given in Case~1:
\begin{gather}
\label{laxB1_kuper}
\chi_{[1]}={\mathcal W}\chi,
\qquad
{\mathcal W}= \left(
\begin{matrix}
\frac{1}{2}v_1 & 1 & 0
\vspace{1mm}\\ \lambda-p_1+\frac{1}{4}{v_1}^2+\xi\eta_{[1]}& \frac{1}{2}v_1 & \xi
\vspace{1mm}\\ \eta_{[1]} & 0 & 1
\end{matrix}
\right),
\\
\label{laxB2_kuper}
\chi_{[2]}={\mathcal N}\chi,
\qquad
{\mathcal N}= \left(
\begin{matrix} \frac{1}{2}v_2 & 1 & 0
\vspace{1mm}\\  \lambda-p_2+\frac{1}{4}{v_2}^2+\xi\eta_{[2]}& \frac{1}{2}v_2 & \xi
\vspace{1mm}\\  \eta_{[2]} & 0 & 1
\end{matrix}
\right).
\end{gather}
Now the compatibility condition of~\eqref{laxB1_kuper} and~\eqref{laxB2_kuper}, namely
\begin{gather*}
{\mathcal W}_{[2]}{\mathcal N}={\mathcal N}_{[1]}{\mathcal W}
\end{gather*}
yields an integrable dif\/ference-dif\/ference system
\begin{gather}
\xi_{[2]}=\xi_{[1]}+\frac{2(p_2-p_1)}{w_{[12]}-w}\xi,
\qquad \eta_{[2]}=\eta_{[1]}+\frac{2(p_1-p_2)}{w_{[12]}-w}\eta_{[12]},
\nonumber\\
 w_{[2]}=w_{[1]}+\frac{4(p_2-p_1)}{w_{[12]}-w}+\frac{8(p_2-p_1)}{(w_{[12]}-w)^2}\xi\eta_{[12]}.\label{dde2}
\end{gather}

{\bf  Case B.} Consider two Darboux transformations which are obtained in Cases~1 and~2 respectively:
\begin{gather}
\label{laxA1_kuper}
\chi_{[1]}={\mathcal W}\chi,
\qquad
{\mathcal W}= \left(
\begin{matrix}
\frac{1}{2}v_1 & 1 & 0
\vspace{1mm}\\  \lambda-p_1+\frac{1}{4}{v_1}^2+\xi\eta_{[1]}& \frac{1}{2}v_1 & \xi
\vspace{1mm}\\  \eta_{[1]} & 0 & 1
\end{matrix}
\right),
\\
\label{laxA2_kuper}
\chi_{[2]}={\mathcal M}\chi,
\qquad
{\mathcal M}= \left(
\begin{matrix}
\frac{1}{2}v_2 & 1 & -\xi_{[2]}
\vspace{1mm}\\  \lambda- p_{2}+\frac{1}{4}{v_2 }^2 & \frac{1}{2}v_2 & -\frac{1}{2}v_2 \xi_{[2]}
\vspace{1mm}\\  -\frac{1}{2}v_2 \eta & -\eta & \lambda- p_{2} -\xi_{[2]}\eta
\end{matrix}
\right).
\end{gather}
Now the compatibility condition of~\eqref{laxA1_kuper} and~\eqref{laxA2_kuper},
\begin{gather*}
{\mathcal W}_{[2]}{\mathcal M}={\mathcal M}_{[1]}{\mathcal W},
\qquad
\end{gather*}
provides us the following integrable dif\/ference-dif\/ference system
\begin{gather}
\xi_{[12]}=\xi+\frac{2(p_1-p_2)}{w_{[1]}-w_{[2]}}\xi_{[2]},
\qquad
\eta_{[21]}=\eta+\frac{2(p_2-p_1)}{w_{[1]}-w_{[2]}}\eta_{[1]},
\nonumber\\
w_{[12]}=w_{[21]}=w+\frac{4(p_1-p_2)}{w_{[1]}-w_{[2]}}+\frac{8(p_1-p_2)}{(w_{[1]}-w_{[2]})^2}\xi_{[2]}\eta_{[1]}.
\label{w12} %\label{dde}
\end{gather}

While for the bosonic f\/ield we have $w_{[12]}=w_{[21]}$, it is not clear from the equations~\eqref{w12} that whether the
same situation appears for the fermionic variables $\xi$ and $\eta$.
We now show that this is indeed the case.
By means of the B\"acklund transformations
\begin{gather*}
{\eta_{[1]}}_x=-\eta+\frac{1}{2} (w_{[1]}-w )\eta_{[1]},
\qquad
\xi_{[21]}={\xi_{[2]}}_x+\frac{1}{2} (w_{[21]}-w_{[2]} )\xi_{[2]},
\\
{\xi_{[2]}}_x=\xi+\frac{1}{2} (w_{[2]}-w )\xi_{[2]},
\qquad
\eta_{[12]}=-{\eta_{[1]}}_x-\frac{1}{2} (w_{[12]}-w_{[1]} )\eta_{[1]},
\end{gather*}
it follows that
\begin{gather*}
\xi_{[21]}= \xi+\frac{1}{2} (w_{[21]}-w )\xi_{[2]},
\qquad
\eta_{[12]}=\eta -\frac{1}{2} (w_{[12]}-w )\eta_{[1]},
\end{gather*}
these equations, taking~\eqref{w12} into account, yield
\begin{gather*}
\xi_{[21]}=\xi+\frac{2(p_1-p_2)}{w_{[1]}-w_{[2]}}\xi_{[2]},
\qquad
\eta_{[12]}=\eta-\frac{2(p_1-p_2)}{w_{[1]}-w_{[2]}}\eta_{[1]},
\end{gather*}
thus
\begin{gather*}
\xi_{[21]}=\xi_{[12]},
\qquad
\eta_{[12]}=\eta_{[21]}.
\end{gather*}

\begin{Remark}
For all fermionic f\/ields vanish, both~\eqref{dde2} and~\eqref{w12} reduce to
\begin{gather*}
w_{[12]}=w+\frac{ 4(p_1- p_2)}{w_{[1]}-w_{[2]}},
\end{gather*}
which is the potential KdV lattice or H1 in Adler--Bobenko--Suris's classif\/ication~\cite{abs} or the classical nonlinear
superposition formula for the KdV equation.
\end{Remark}

Finally we consider

{\bf Case C.} Assume a~pair of Darboux transformations which are provided in Case~3:
\begin{gather}
\label{lax_Binary_kuper}
\chi_{[1]}={\mathcal W}\chi,
\\
\chi_{[2]}={\mathcal V}\chi,
\label{lax2_Binary_kuper}
\end{gather}
where
\begin{gather*}
{\mathcal W}= \left(
\begin{matrix} \lambda+A& \frac{1}{2} v_1 & \xi -\xi_{[1]}
\vspace{1mm}\\ \frac{1}{2}\lambda v_1+B & \lambda+A +\frac{{v_1}_x}{2} & -\left(\frac{v_1}{4}+\frac{{v_1}_x}{2 v_1}\right)(\xi_{[1]}-\xi)
\vspace{1mm}\\  \left(\frac{v_1}{4}-\frac{{v_1}_x}{2 v_1}\right)(\eta_{[1]}-\eta)& \eta_{[1]}-\eta & \lambda-p_1+\frac{2}{v_1}(\xi_{[1]}-\xi)(\eta_{[1]}-\eta)
\end{matrix}
\right)
\end{gather*}
with
\begin{gather*}
A\equiv -p_1-\frac{{v_1}_x}{4}+\frac{{v_1}^2}{8}+\frac{1}{v_1} (\xi_{[1]}-\xi ) (\eta_{[1]}-\eta ),
\\
B\equiv-\frac{{{v_1}_{x}}^2}{8 v_1}+\frac{{v_1}^3}{32}-\frac{p_1 }{2} v_1 + \frac{1}{2}
 (\xi_{[1]}-\xi ) (\eta_{[1]}-\eta ),
\end{gather*}
the matrix ${\mathcal V}$ is the matrix ${\mathcal W}$ with $ p_1$, $\xi_{[1]}$, $\eta_{[1]}$ and $w_{[1]}$ replaced by $
p_2$, $\xi_{[2]}$, $\eta_{[2]}$ and $w_{[2]}$ respectively.
By the compatibility $(\chi_{[1]})_{[2]}=(\chi_{[2]})_{[1]}$ of~\eqref{lax_Binary_kuper} and~\eqref{lax2_Binary_kuper},
i.e.\
the Bianchi permutability of the B\"acklund transformation~\eqref{eq9} with $\xi_{[21]}=\xi_{[12]}$,
$\eta_{[21]}=\eta_{[12]}$ and $w_{[21]}=w_{[12]}$, we f\/ind that the following consistency condition
\begin{gather}
\label{eq2+}
{\mathcal W}_{[2]}{\mathcal V}={\mathcal V}_{[1]}{\mathcal W}
\end{gather}
must be true. \eqref{eq2+}~leads to
\begin{gather*}%\label{condition1}
{w_{[12]}}_x={w_{[2]}}_x+\frac{1}{2} (w_{[12]}-w_{[2]} )^2
+ (w_{[12]}-w_{[2]} )\left[\frac{1}{2} (w_{[2]}-w_{[1]} )
+ (\ln (w_{[1]}-w_{[2]} ) )_x\right]
\\
\phantom{{w_{[12]}}_x=}
{}-(w_{[1]}-w_{[2]})^{-1} \left[\frac{1}{2} h+4c (w_{[12]}-w_{[2]}-w_{[1]}+w )\right]
\\
\phantom{{w_{[12]}}_x=}
{}-4 \big[ v_1v_2 (w_{[12]}-w_{[2]} ) (w_{[1]}-w_{[2]} ) (w_{[12]}-w_{[1]} ) \big]^{-1}
\\
\phantom{{w_{[12]}}_x=}{}
\times
\Big\{ -v_2 (w_{[12]}-w_{[2]} ) (w_{[12]}-w )^2 (\xi_{[1]}-\xi)(\eta_{[1]}-\eta)
\\
\phantom{{w_{[12]}}_x=}
{}-v_1v_2 (w_{[12]}-w ) (w_{[1]}-w_{[2]} )(\xi_{[12]}-\xi_{[2]})(\eta_{[12]}-\eta_{[2]})
\\
\phantom{{w_{[12]}}_x=}
{}+v_1 (w_{[12]}-w_{[2]} ) (w_{[12]}-w ) (w_{[12]}-w_{[2]}-v_1 )(\xi_{[2]}-\xi)(\eta_{[2]}-\eta)
\\
\phantom{{w_{[12]}}_x=}
{}+v_{1}^2v_2 (w_{[12]}-w_{[2]} )\big[(\xi_{[1]}-\xi)(\eta_{[12]}-\eta)-(\xi_{[2]}-\xi)(\eta_{[12]}-\eta_{[2]})\big]
\\
\phantom{{w_{[12]}}_x=}
{}+v_1v_2 (w_{[12]}-w_{[2]} ) (w_{[1]}+w-2w_{[12]} )
\\
\phantom{{w_{[12]}}_x=}{}
\times\big[(\eta_{[1]}-\eta)(\xi_{[12]}-\xi)-(\eta_{[2]}-\eta)(\xi_{[12]}-\xi_{[2]})\big]\Big\},
\end{gather*}
and
\begin{gather}
\xi_{[12]}=\xi+f_1(\xi_{1}-\xi)+f_2(\xi_{2}-\xi) + (\xi_{1}-\xi)(\xi_{2}-\xi) [f_3(\eta_{1}-\eta)+f_4
(\eta_{2}-\eta) ],
\nonumber\\
\eta_{[12]}=\eta+f_1(\eta_{1}-\eta)+f_2(\eta_{2}-\eta) +
(\eta_{1}-\eta)(\eta_{2}-\eta)[f_3(\xi_{1}-\xi)+f_4(\xi_{2}-\xi)],
\nonumber\\
w_{[12]}=w +g_1 +g_2[(\xi_{1}-\xi)(\eta_{2}-\eta)+(\eta_{1}-\eta)(\xi_{2}-\xi)]
\nonumber
\\
\hphantom{w_{[12]}=}{}
+g_3 (\xi_{1}-\xi)(\xi_{2}-\xi)(\eta_{1}-\eta)(\eta_{2}-\eta),
\label{eq26}
\end{gather}
where
\begin{gather*}
c\equiv p_1-p_2,
\qquad
h\equiv v_1 v_2(v_1-v_2) +2(v_1 {v_{2}}_x-v_2 {v_{1}}_x),
\\
f_1\equiv \frac{8 c v_2 (h-8 c v_1)}{h^2-64 c^2 v_1v_2},
\qquad
f_2\equiv \frac{8 c v_1 (h-8 c v_2)}{h^2-64 c^2 v_1v_2},
\\
f_3\equiv \frac{-64 c v_2} {(h^2-64 c^2 v_1v_2)^2}\left[h^2-16 c h v_1+64 c^2 v_1v_2 \right],
\\
f_4\equiv \frac{-64 c v_1} {(h^2-64 c^2 v_1v_2)^2}\left[h^2-16 c h v_2+64 c^2 v_1v_2 \right],
\\
g_1\equiv \frac{16 c v_1v_2 } {h^2-64 c^2 v_1v_2} \left[h-4 c(v_1+v_2) \right],
\qquad
g_2\equiv \frac{128 c v_1v_2} {(h^2-64 c^2 v_1v_2)^2} \left[h-8 c v_1\right]\left[h-8 c v_2 \right],
\\
g_3\equiv \frac{2048 c v_1v_2}{(h^2-64 c^2 v_1v_2)^3} \left[-h^3+12 c h^2 (v_1+v_2)-192 c^2 h v_1v_2 +256 c^3v_1v_2
(v_1+v_2)\right].
\end{gather*}
Here the system~\eqref{eq26} serves as a~discrete system which, being cumbersome, nevertheless has the advantage that it
is easy to handle if one considers the reductions.
In fact, we have two interesting cases:

1.~For $\xi=\eta$, the consistency condition~\eqref{eq2+} leads to
\begin{gather}
{w_{[12]}}_x={w_{[2]}}_x+\frac{1}{2} (w_{[12]}-w_{[2]} )^2
+ (w_{[12]}-w_{[2]} )\left[\frac{1}{2} (w_{[2]}-w_{[1]} )+ (\ln(w_{[1]}-w_{[2]}) )_x\right]
\nonumber
\\
\phantom{{w_{[12]}}_x=}
{}- \frac{h+8c (w_{[12]}-w_{[2]}-w_{[1]}+w )+16 (\xi_{[12]}-\xi ) (\xi_{[1]}-\xi_{[2]} )}{2 ({w_{[1]}-w_{[2]}} )},
\label{condition2}
\end{gather}
and
\begin{gather}
\xi_{[12]}=\xi+f_1(\xi_{1}-\xi)+f_2(\xi_{2}-\xi),
\qquad
w_{[12]}=w +g_1+2g_2(\xi_{1}-\xi)(\xi_{2}-\xi).\label{eq28}
\end{gather}
By a~direct calculation, one can check that~\eqref{condition2} is satisf\/ied if we substitute~\eqref{eq28} into it and
take account of the corresponding B\"acklund transformations~\eqref{eq15} into consideration.

2.~For all fermionic f\/ields vanish,~\eqref{eq2+} leads to
\begin{gather}
{w_{[12]}}_x={w_{[2]}}_x + (w_{[12]}-w_{[2]} )\left[\frac{1}{2} (w_{[2]}-w_{[1]} )
+ (\ln (w_{[1]}-w_{[2]} ) )_x\right]
\nonumber
\\
\phantom{{w_{[12]}}_x=}
{}+\frac{1}{2} (w_{[12]}-w_{[2]} )^2- \frac{h+8c (w_{[12]}-w_{[2]}-w_{[1]}+w )}{2 ({w_{[1]}-w_{[2]}} )},\label{condition3}
\end{gather}
and
\begin{gather}
\label{NSF_kdv}
w_{[12]}=w +g_1.
\end{gather}
Also, a~direct calculation shows that~\eqref{condition3} is satisf\/ied if we substitute~\eqref{NSF_kdv} into it and take
the corresponding B\"acklund transformations~\eqref{levi_bt} into consideration.
Thus~\eqref{NSF_kdv} may be regarded as the nonlinear superposition formula for the B\"acklund
transformation~\eqref{levi_bt} for the KdV equation.
To the best of our knowledge, this result is also new.

\section{Conclusion}

In this paper, we have constructed three types of B\"acklund and Darboux transformations for a~generalized super KdV
equation.
By means of these transformations, the super KdV equation has been discretized.
In particular, by considering the reductions, we have succeeded in obtaining a~B\"{a}cklund transformation and a~discrete
version for Kupershmidt's super KdV equation.
As a~by-product, we have found a~nonlinear superposition formula for the B\"acklund transformation obtained by Levi
early.
The discretization for the supersymmetric Schr\"{o}dinger equation~\cite{snls} is under investigated and will appear
elsewhere.

\subsection*{Acknowledgments}

The comments of the anonymous referees have been useful in clarifying certain points such as the connection between
Levi's B\"{a}cklund transformation and elementary B\"{a}cklund transformation of the KdV equation.
This work is supported by the National Natural Science Foundation of China (grant numbers: 10971222, 11271366 and
11331008) and the Fundamental Research Funds for Central Universities.

\pdfbookmark[1]{References}{ref}
\LastPageEnding

\end{document}